\documentstyle[aps,multicol]{revtex}

\begin{document}

\draft

\title{Bosons with attractive interactions in a trap: 
Is the ground state fragmented ?}

\author{\O.\ Elgar\o y$^a$ and C.\ J.\ Pethick $^b$}

\address{$^a$Department of Physics, University of Oslo, N-0316 Oslo, Norway}

\address{$^b$Nordita, Blegdamsvej 17, DK-2100 Copenhagen \O, Denmark}

\maketitle

\begin{abstract}
Possible fragmentation of a Bose-Einstein condensate with negative 
scattering length is investigated using a simple two-level model.  
Our results indicate that 
fragmentation does not take place for values of the 
coupling for which the system is metastable.  We also comment 
on the possibility of realizing a fragmented condensate in 
trapping potentials other than an harmonic one. 
\end{abstract}


\begin{multicols}{2}

Bose-Einstein condensation in dilute atomic gases has been 
an area of intense activity since the first experimental 
realizations of these systems \cite{and95,davis95,brad95}.  
The experiments have so far been carried out with the 
alkali atoms $^{87}{\rm Rb}$, $^{23}{\rm Na}$, and 
$^7{\rm Li}$.  While the s-wave scattering length is positive  
in the first two cases, it is negative for $^7{\rm Li}$, 
implying that the effective atom-atom interaction is 
attractive.  A homogeneous system of these atoms will  
collapse to a dense state before densities where 
a Bose-Einstein condensate can form are reached \cite{stoof94}. 
However, in the experiments 
Bose condensation is realized under inhomogeneous conditions, 
and theoretical calculations show that the system may then be 
metastable as long as the number of condensed particles is 
below a critical value, $N_c\sim 10^3$ for $^7{\rm Li}$ 
\cite{houb96,rupr95} for the trap used in the experiments.  
The latest experiments are consistent with this result \cite{brad97}.  

The homogeneous Bose gas with attractive interactions 
was considered in a 
paper by Nozi\`{e}res and Saint James \cite{noz82}.  
While their main interest was the onset of a 
BCS-like transition, they also 
asked whether a fragmented Bose condensate would 
form.  In the homogeneous case, fragmentation means that 
the particle distribution in momentum space, $n_{{\bf k}}$ has 
a sharp peak near ${\bf k}=0$ which extends over a number of 
$({\bf k},-{\bf k})$ states large compared to 1, but small compared 
to the particle number $N$: on a macroscopic scale, it 
looks like a $\delta$-function.  
That this is a reasonable question can be seen within the 
Hartree-Fock approximation: for bosons with an attractive 
interaction the Fock term makes it energetically 
favorable to spread the particles over several states.  
Still, the authors of Ref. \cite{noz82} found that 
fragmentation of the condensate does not take place in 
a homogeneous system.  However, the physics of 
the inhomogeneous Bose gas is different, and it seems worthwhile 
to ask the same question again in this context.  

We want to study the ground state of bosons with a 
negative s-wave scattering length in an isotropic 
harmonic oscillator potential. 
Since the question we are addressing is whether the condensate 
can fragment, we must allow for (at least) two single-particle 
states in the calculation.  
We will consider the two normalized single-particle wave functions 
\begin{equation}
 \psi_0({\bf r})=\frac{1}{\pi^{3/4}b^{3/2}}e^{-r^2/2b^2} 
\label{eq:eq1}
\end{equation}
\begin{equation}
 \psi_1({\bf r})=\frac{2^{1/2}}{\pi^{3/4}b^{5/2}}ze^{-r^2/2b^2} 
\label{eq:eq2}
\end{equation}
where $r^2=x^2+y^2+z^2$ and $b$ is a variational parameter.  
These have the same shape as the ground state and the first excited 
state with angular momentum projection $m_{l}=0$ 
of the harmonic oscillator well.  There will be interactions 
between particles in the different levels and also between 
particles in the same level.  For fixed $b$ the many-body 
Hamiltonian is 
\begin{eqnarray}
\hat{H}&=& \sum_i \langle i|\hat{t}+\hat{u}|i\rangle \hat{a}_i^{\dagger}
\hat{a}_i \nonumber \\
 &+&\frac{1}{2}\sum_{ijkl}V_{ijkl}\hat{a}_i^{\dagger}\hat{a}_j^{\dagger}
\hat{a}_k \hat{a}_l 
\label{eq:eq3}
\end{eqnarray}
where $i,j,k,l=0,1$, $\hat{t}$ is the kinetic energy operator, 
$\hat{u}=\frac{1}{2}m\omega^2 r^2$ and 
\begin{equation}
V_{ijkl}=-\frac{4\pi\hbar^2|a|}{m}\int d^3 r \psi_i^*\psi_j^*
\psi_k\psi_l,
\label{eq:eq4}
\end{equation}
where $a$ is the scattering length (assumed negative) and $m$ is the 
boson mass.  We will in the following   
solve this two-level problem for arbitrary values of $b$, thus obtaining 
the ground state energy of the model as a function of this 
variational parameter.  In the next step $b$ is determined by minimizing 
the energy.     
This simple two-level system can be solved exactly 
and is in fact a special case of the so-called Lipkin model 
\cite{lip65} which has been widely used in nuclear physics as 
a tool for studying many-body approximation methods.   

\begin{figure}
\setlength{\unitlength}{0.1bp}
\special{!
/gnudict 40 dict def
gnudict begin
/Color false def
/Solid false def
/gnulinewidth 5.000 def
/vshift -33 def
/dl {10 mul} def
/hpt 31.5 def
/vpt 31.5 def
/M {moveto} bind def
/L {lineto} bind def
/R {rmoveto} bind def
/V {rlineto} bind def
/vpt2 vpt 2 mul def
/hpt2 hpt 2 mul def
/Lshow { currentpoint stroke M
  0 vshift R show } def
/Rshow { currentpoint stroke M
  dup stringwidth pop neg vshift R show } def
/Cshow { currentpoint stroke M
  dup stringwidth pop -2 div vshift R show } def
/DL { Color {setrgbcolor Solid {pop []} if 0 setdash }
 {pop pop pop Solid {pop []} if 0 setdash} ifelse } def
/BL { stroke gnulinewidth 2 mul setlinewidth } def
/AL { stroke gnulinewidth 2 div setlinewidth } def
/PL { stroke gnulinewidth setlinewidth } def
/LTb { BL [] 0 0 0 DL } def
/LTa { AL [1 dl 2 dl] 0 setdash 0 0 0 setrgbcolor } def
/LT0 { PL [] 0 1 0 DL } def
/LT1 { PL [4 dl 2 dl] 0 0 1 DL } def
/LT2 { PL [2 dl 3 dl] 1 0 0 DL } def
/LT3 { PL [1 dl 1.5 dl] 1 0 1 DL } def
/LT4 { PL [5 dl 2 dl 1 dl 2 dl] 0 1 1 DL } def
/LT5 { PL [4 dl 3 dl 1 dl 3 dl] 1 1 0 DL } def
/LT6 { PL [2 dl 2 dl 2 dl 4 dl] 0 0 0 DL } def
/LT7 { PL [2 dl 2 dl 2 dl 2 dl 2 dl 4 dl] 1 0.3 0 DL } def
/LT8 { PL [2 dl 2 dl 2 dl 2 dl 2 dl 2 dl 2 dl 4 dl] 0.5 0.5 0.5 DL } def
/P { stroke [] 0 setdash
  currentlinewidth 2 div sub M
  0 currentlinewidth V stroke } def
/D { stroke [] 0 setdash 2 copy vpt add M
  hpt neg vpt neg V hpt vpt neg V
  hpt vpt V hpt neg vpt V closepath stroke
  P } def
/A { stroke [] 0 setdash vpt sub M 0 vpt2 V
  currentpoint stroke M
  hpt neg vpt neg R hpt2 0 V stroke
  } def
/B { stroke [] 0 setdash 2 copy exch hpt sub exch vpt add M
  0 vpt2 neg V hpt2 0 V 0 vpt2 V
  hpt2 neg 0 V closepath stroke
  P } def
/C { stroke [] 0 setdash exch hpt sub exch vpt add M
  hpt2 vpt2 neg V currentpoint stroke M
  hpt2 neg 0 R hpt2 vpt2 V stroke } def
/T { stroke [] 0 setdash 2 copy vpt 1.12 mul add M
  hpt neg vpt -1.62 mul V
  hpt 2 mul 0 V
  hpt neg vpt 1.62 mul V closepath stroke
  P  } def
/S { 2 copy A C} def
end
}
\begin{picture}(3600,2160)(0,0)
\special{"
gnudict begin
gsave
50 50 translate
0.100 0.100 scale
0 setgray
/Helvetica findfont 100 scalefont setfont
newpath
-500.000000 -500.000000 translate
LTa
600 870 M
2817 0 V
600 251 M
0 1858 V
LTb
600 251 M
63 0 V
2754 0 R
-63 0 V
600 561 M
63 0 V
2754 0 R
-63 0 V
600 870 M
63 0 V
2754 0 R
-63 0 V
600 1180 M
63 0 V
2754 0 R
-63 0 V
600 1490 M
63 0 V
2754 0 R
-63 0 V
600 1799 M
63 0 V
2754 0 R
-63 0 V
600 2109 M
63 0 V
2754 0 R
-63 0 V
600 251 M
0 63 V
0 1795 R
0 -63 V
976 251 M
0 63 V
0 1795 R
0 -63 V
1351 251 M
0 63 V
0 1795 R
0 -63 V
1727 251 M
0 63 V
0 1795 R
0 -63 V
2102 251 M
0 63 V
0 1795 R
0 -63 V
2478 251 M
0 63 V
0 1795 R
0 -63 V
2854 251 M
0 63 V
0 1795 R
0 -63 V
3229 251 M
0 63 V
0 1795 R
0 -63 V
600 251 M
2817 0 V
0 1858 V
-2817 0 V
600 251 L
LT0
3114 1946 M
180 0 V
976 1486 M
1 4 V
3 8 V
4 10 V
5 11 V
6 13 V
8 12 V
8 10 V
10 9 V
11 5 V
12 3 V
13 -1 V
14 -5 V
16 -7 V
16 -11 V
18 -14 V
18 -15 V
20 -18 V
20 -20 V
22 -20 V
23 -22 V
23 -21 V
24 -22 V
26 -22 V
26 -21 V
27 -21 V
28 -20 V
29 -20 V
29 -18 V
31 -18 V
30 -16 V
32 -16 V
32 -15 V
33 -13 V
34 -13 V
34 -12 V
34 -11 V
35 -10 V
36 -9 V
36 -8 V
36 -8 V
37 -6 V
37 -7 V
37 -5 V
37 -5 V
38 -4 V
38 -3 V
38 -3 V
38 -3 V
38 -2 V
38 -1 V
39 -1 V
38 -1 V
38 -1 V
37 0 V
38 1 V
38 1 V
37 0 V
37 2 V
36 1 V
37 2 V
36 2 V
35 2 V
35 2 V
35 3 V
34 2 V
33 3 V
33 3 V
33 3 V
31 3 V
31 3 V
30 3 V
30 3 V
29 3 V
27 3 V
28 3 V
26 3 V
25 4 V
25 3 V
23 3 V
23 3 V
21 2 V
21 3 V
19 3 V
19 2 V
18 3 V
16 2 V
15 3 V
15 2 V
13 2 V
12 1 V
11 2 V
9 1 V
9 2 V
7 1 V
7 1 V
5 1 V
3 0 V
3 1 V
2 0 V
LT1
3114 1846 M
180 0 V
1057 1578 M
14 -6 V
16 -8 V
16 -12 V
18 -14 V
18 -17 V
20 -18 V
20 -20 V
22 -21 V
23 -22 V
23 -22 V
24 -22 V
26 -22 V
26 -21 V
27 -21 V
28 -21 V
29 -19 V
29 -19 V
31 -17 V
30 -17 V
32 -16 V
32 -14 V
33 -14 V
34 -13 V
34 -12 V
34 -11 V
35 -10 V
36 -9 V
36 -8 V
36 -8 V
37 -6 V
37 -6 V
37 -6 V
37 -5 V
38 -4 V
38 -3 V
38 -3 V
38 -3 V
38 -2 V
38 -1 V
39 -1 V
38 -1 V
38 -1 V
37 1 V
38 0 V
38 1 V
37 0 V
37 2 V
36 1 V
37 2 V
36 2 V
35 2 V
35 2 V
35 3 V
34 2 V
33 3 V
33 3 V
33 3 V
31 3 V
31 3 V
30 3 V
30 3 V
29 3 V
27 3 V
28 3 V
26 3 V
25 4 V
25 3 V
23 3 V
23 3 V
21 2 V
21 3 V
19 3 V
19 2 V
18 3 V
16 2 V
15 3 V
15 2 V
13 2 V
12 1 V
11 2 V
9 1 V
9 2 V
7 1 V
7 1 V
5 1 V
3 0 V
3 1 V
2 0 V
LT2
3114 1746 M
180 0 V
1099 251 M
5 39 V
9 66 V
11 66 V
11 65 V
13 62 V
14 58 V
14 55 V
16 54 V
17 49 V
18 43 V
19 38 V
19 32 V
21 31 V
22 27 V
23 22 V
23 17 V
24 15 V
26 14 V
26 10 V
27 8 V
27 6 V
28 6 V
30 4 V
29 3 V
31 2 V
31 1 V
31 1 V
33 1 V
32 0 V
33 0 V
34 0 V
34 0 V
35 0 V
35 0 V
35 0 V
36 0 V
35 0 V
36 0 V
37 0 V
36 1 V
37 1 V
36 1 V
37 1 V
37 1 V
36 1 V
37 2 V
36 2 V
37 2 V
36 2 V
36 2 V
36 3 V
35 2 V
36 3 V
35 3 V
34 2 V
34 3 V
34 4 V
33 3 V
33 3 V
32 3 V
32 4 V
31 3 V
30 4 V
30 3 V
29 3 V
28 4 V
28 3 V
27 4 V
26 3 V
25 4 V
24 3 V
24 3 V
23 3 V
21 3 V
21 3 V
20 3 V
19 3 V
18 3 V
16 2 V
16 2 V
15 3 V
14 2 V
12 2 V
12 2 V
10 1 V
10 2 V
8 1 V
7 1 V
6 1 V
5 1 V
4 1 V
2 0 V
2 0 V
LT3
3114 1646 M
180 0 V
1094 251 M
9 71 V
18 113 V
18 99 V
20 88 V
20 76 V
22 65 V
23 56 V
23 47 V
24 39 V
26 33 V
26 26 V
27 22 V
28 17 V
29 14 V
29 10 V
31 8 V
30 6 V
32 4 V
32 -3 V
33 2 V
34 1 V
34 1 V
34 1 V
35 0 V
36 0 V
36 0 V
36 -1 V
37 0 V
37 0 V
37 0 V
37 0 V
38 0 V
38 0 V
38 1 V
38 1 V
38 0 V
38 1 V
39 2 V
38 1 V
38 2 V
37 2 V
38 2 V
38 2 V
37 2 V
37 3 V
36 2 V
37 3 V
36 3 V
35 3 V
35 3 V
35 3 V
34 4 V
33 3 V
33 3 V
33 4 V
31 3 V
31 4 V
30 3 V
30 4 V
29 3 V
27 4 V
28 3 V
26 4 V
25 3 V
25 4 V
23 3 V
23 3 V
21 3 V
21 3 V
19 3 V
19 3 V
18 2 V
16 3 V
15 2 V
15 2 V
13 2 V
12 2 V
11 2 V
9 1 V
9 2 V
7 1 V
7 1 V
5 1 V
3 0 V
3 1 V
2 0 V
LT4
3114 1546 M
180 0 V
1239 251 M
8 39 V
24 102 V
26 88 V
26 76 V
27 66 V
28 57 V
29 49 V
29 41 V
31 36 V
30 30 V
32 26 V
32 22 V
33 19 V
34 16 V
34 13 V
34 12 V
35 10 V
36 8 V
36 8 V
36 6 V
37 6 V
37 5 V
37 5 V
37 5 V
38 4 V
38 4 V
38 3 V
38 4 V
38 3 V
38 4 V
39 3 V
38 4 V
38 3 V
37 3 V
38 4 V
38 4 V
37 3 V
37 4 V
36 3 V
37 4 V
36 4 V
35 4 V
35 4 V
35 4 V
34 3 V
33 4 V
33 4 V
33 4 V
31 4 V
31 4 V
30 4 V
30 4 V
29 4 V
27 4 V
28 4 V
26 3 V
25 4 V
25 4 V
23 3 V
23 3 V
21 4 V
21 3 V
19 3 V
19 3 V
18 2 V
16 3 V
15 3 V
15 2 V
13 2 V
12 2 V
11 2 V
9 1 V
9 2 V
7 1 V
7 1 V
5 1 V
3 1 V
3 0 V
2 0 V
LT5
3114 1446 M
180 0 V
1234 251 M
13 64 V
24 100 V
26 86 V
26 74 V
27 64 V
28 56 V
29 47 V
29 40 V
31 35 V
30 29 V
32 25 V
32 21 V
33 19 V
34 15 V
34 13 V
34 11 V
35 10 V
36 8 V
36 7 V
36 6 V
37 1 V
37 5 V
37 5 V
37 4 V
38 4 V
38 4 V
38 4 V
38 3 V
38 4 V
38 3 V
39 3 V
38 4 V
38 3 V
37 4 V
38 3 V
38 4 V
37 3 V
37 4 V
36 4 V
37 3 V
36 4 V
35 4 V
35 4 V
35 4 V
34 4 V
33 3 V
33 4 V
33 4 V
31 4 V
31 4 V
30 4 V
30 4 V
29 4 V
27 4 V
28 4 V
26 3 V
25 4 V
25 4 V
23 3 V
23 3 V
21 4 V
21 3 V
19 3 V
19 3 V
18 3 V
16 2 V
15 3 V
15 2 V
13 2 V
12 2 V
11 2 V
9 1 V
9 2 V
7 1 V
7 1 V
5 1 V
3 1 V
3 0 V
2 0 V
stroke
grestore
end
showpage
}
\put(3054,1446){\makebox(0,0)[r]{Semi-classical, $N|a|/a_{{\rm osc}}=1.0$}}
\put(3054,1546){\makebox(0,0)[r]{Exact, $N|a|/a_{{\rm osc}}=1.0$}}
\put(3054,1646){\makebox(0,0)[r]{Semi-classical, $N|a|/a_{{\rm osc}}=0.67$ }}
\put(3054,1746){\makebox(0,0)[r]{Exact, $N|a|/a_{{\rm osc}}=0.67$}}
\put(3054,1846){\makebox(0,0)[r]{Semi-classical, $N|a|/a_{{\rm osc}}=0.3$}}
\put(3054,1946){\makebox(0,0)[r]{Exact, $N|a|/a_{{\rm osc}}=0.3$}}
\put(2008,51){\makebox(0,0){$x=b/a_{{\rm osc}}$}}
\put(100,1180){%
\special{ps: gsave currentpoint currentpoint translate
270 rotate neg exch neg exch translate}%
\makebox(0,0)[b]{\shortstack{$E/N\hbar\omega$}}%
\special{ps: currentpoint grestore moveto}%
}
\put(3229,151){\makebox(0,0){1.4}}
\put(2854,151){\makebox(0,0){1.2}}
\put(2478,151){\makebox(0,0){1}}
\put(2102,151){\makebox(0,0){0.8}}
\put(1727,151){\makebox(0,0){0.6}}
\put(1351,151){\makebox(0,0){0.4}}
\put(976,151){\makebox(0,0){0.2}}
\put(600,151){\makebox(0,0){0}}
\put(540,2109){\makebox(0,0)[r]{8}}
\put(540,1799){\makebox(0,0)[r]{6}}
\put(540,1490){\makebox(0,0)[r]{4}}
\put(540,1180){\makebox(0,0)[r]{2}}
\put(540,870){\makebox(0,0)[r]{0}}
\put(540,561){\makebox(0,0)[r]{-2}}
\put(540,251){\makebox(0,0)[r]{-4}}
\end{picture}
	\caption{Comparison of the exact and semi-classical result 
for the energy per particle.}
    \label{fig:fig1}
\end{figure}
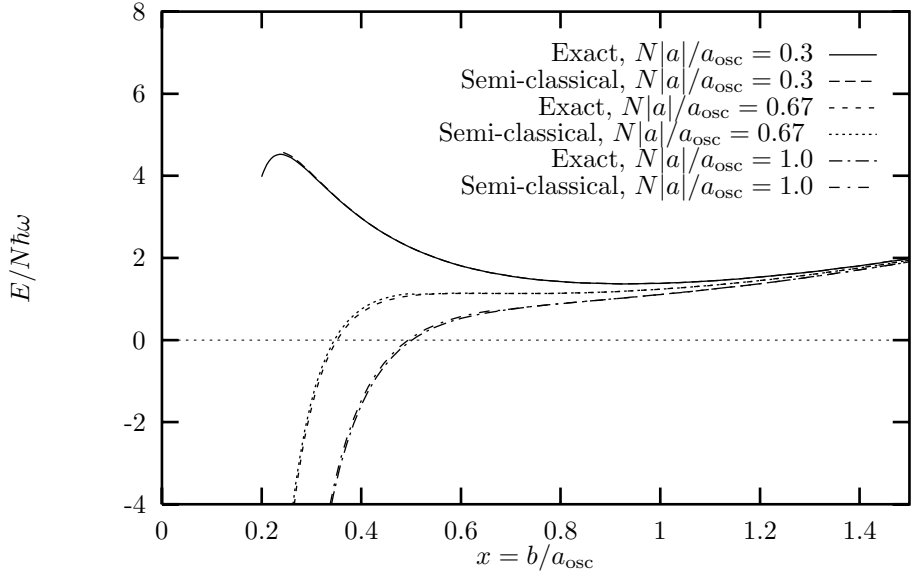

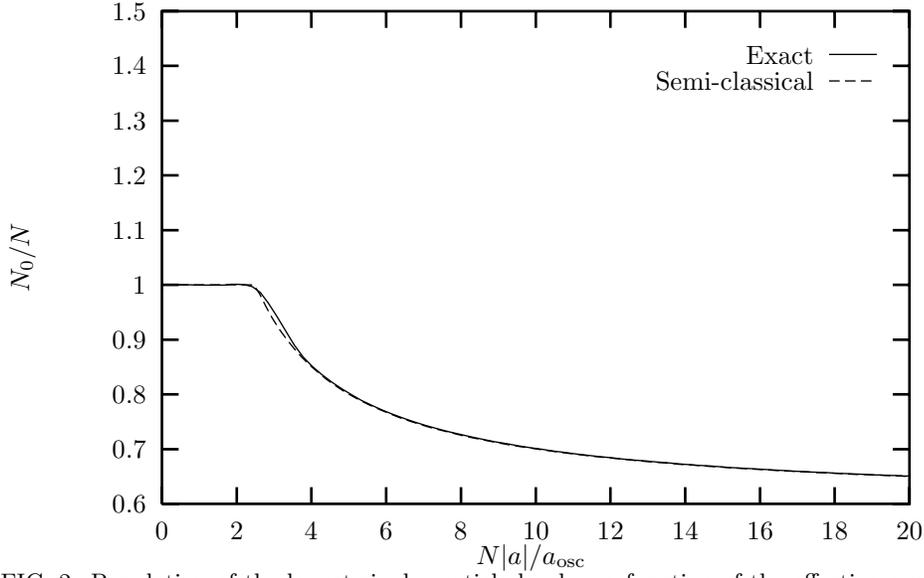
\begin{figure}
\setlength{\unitlength}{0.1bp}
\special{!
/gnudict 40 dict def
gnudict begin
/Color false def
/Solid false def
/gnulinewidth 5.000 def
/vshift -33 def
/dl {10 mul} def
/hpt 31.5 def
/vpt 31.5 def
/M {moveto} bind def
/L {lineto} bind def
/R {rmoveto} bind def
/V {rlineto} bind def
/vpt2 vpt 2 mul def
/hpt2 hpt 2 mul def
/Lshow { currentpoint stroke M
  0 vshift R show } def
/Rshow { currentpoint stroke M
  dup stringwidth pop neg vshift R show } def
/Cshow { currentpoint stroke M
  dup stringwidth pop -2 div vshift R show } def
/DL { Color {setrgbcolor Solid {pop []} if 0 setdash }
 {pop pop pop Solid {pop []} if 0 setdash} ifelse } def
/BL { stroke gnulinewidth 2 mul setlinewidth } def
/AL { stroke gnulinewidth 2 div setlinewidth } def
/PL { stroke gnulinewidth setlinewidth } def
/LTb { BL [] 0 0 0 DL } def
/LTa { AL [1 dl 2 dl] 0 setdash 0 0 0 setrgbcolor } def
/LT0 { PL [] 0 1 0 DL } def
/LT1 { PL [4 dl 2 dl] 0 0 1 DL } def
/LT2 { PL [2 dl 3 dl] 1 0 0 DL } def
/LT3 { PL [1 dl 1.5 dl] 1 0 1 DL } def
/LT4 { PL [5 dl 2 dl 1 dl 2 dl] 0 1 1 DL } def
/LT5 { PL [4 dl 3 dl 1 dl 3 dl] 1 1 0 DL } def
/LT6 { PL [2 dl 2 dl 2 dl 4 dl] 0 0 0 DL } def
/LT7 { PL [2 dl 2 dl 2 dl 2 dl 2 dl 4 dl] 1 0.3 0 DL } def
/LT8 { PL [2 dl 2 dl 2 dl 2 dl 2 dl 2 dl 2 dl 4 dl] 0.5 0.5 0.5 DL } def
/P { stroke [] 0 setdash
  currentlinewidth 2 div sub M
  0 currentlinewidth V stroke } def
/D { stroke [] 0 setdash 2 copy vpt add M
  hpt neg vpt neg V hpt vpt neg V
  hpt vpt V hpt neg vpt V closepath stroke
  P } def
/A { stroke [] 0 setdash vpt sub M 0 vpt2 V
  currentpoint stroke M
  hpt neg vpt neg R hpt2 0 V stroke
  } def
/B { stroke [] 0 setdash 2 copy exch hpt sub exch vpt add M
  0 vpt2 neg V hpt2 0 V 0 vpt2 V
  hpt2 neg 0 V closepath stroke
  P } def
/C { stroke [] 0 setdash exch hpt sub exch vpt add M
  hpt2 vpt2 neg V currentpoint stroke M
  hpt2 neg 0 R hpt2 vpt2 V stroke } def
/T { stroke [] 0 setdash 2 copy vpt 1.12 mul add M
  hpt neg vpt -1.62 mul V
  hpt 2 mul 0 V
  hpt neg vpt 1.62 mul V closepath stroke
  P  } def
/S { 2 copy A C} def
end
}
\begin{picture}(3600,2160)(0,0)
\special{"
gnudict begin
gsave
50 50 translate
0.100 0.100 scale
0 setgray
/Helvetica findfont 100 scalefont setfont
newpath
-500.000000 -500.000000 translate
LTa
600 251 M
0 1858 V
LTb
600 251 M
63 0 V
2754 0 R
-63 0 V
600 457 M
63 0 V
2754 0 R
-63 0 V
600 664 M
63 0 V
2754 0 R
-63 0 V
600 870 M
63 0 V
2754 0 R
-63 0 V
600 1077 M
63 0 V
2754 0 R
-63 0 V
600 1283 M
63 0 V
2754 0 R
-63 0 V
600 1490 M
63 0 V
2754 0 R
-63 0 V
600 1696 M
63 0 V
2754 0 R
-63 0 V
600 1903 M
63 0 V
2754 0 R
-63 0 V
600 2109 M
63 0 V
2754 0 R
-63 0 V
600 251 M
0 63 V
0 1795 R
0 -63 V
882 251 M
0 63 V
0 1795 R
0 -63 V
1163 251 M
0 63 V
0 1795 R
0 -63 V
1445 251 M
0 63 V
0 1795 R
0 -63 V
1727 251 M
0 63 V
0 1795 R
0 -63 V
2009 251 M
0 63 V
0 1795 R
0 -63 V
2290 251 M
0 63 V
0 1795 R
0 -63 V
2572 251 M
0 63 V
0 1795 R
0 -63 V
2854 251 M
0 63 V
0 1795 R
0 -63 V
3135 251 M
0 63 V
0 1795 R
0 -63 V
3417 251 M
0 63 V
0 1795 R
0 -63 V
600 251 M
2817 0 V
0 1858 V
-2817 0 V
600 251 L
LT0
3114 1946 M
180 0 V
600 1077 M
2 0 V
3 0 V
5 0 V
6 0 V
7 0 V
8 0 V
10 0 V
11 0 V
13 0 V
14 0 V
15 0 V
16 0 V
18 -1 V
19 0 V
20 0 V
22 0 V
22 0 V
24 0 V
25 3 V
26 0 V
27 -2 V
28 -7 V
30 -22 V
30 -40 V
31 -48 V
32 -53 V
34 -56 V
34 -46 V
34 -35 V
36 -31 V
37 -28 V
37 -25 V
38 -23 V
38 -22 V
40 -19 V
40 -18 V
40 -16 V
41 -15 V
41 -14 V
42 -13 V
43 -12 V
42 -11 V
43 -10 V
44 -10 V
43 -9 V
44 -9 V
44 -7 V
43 -8 V
44 -7 V
45 -6 V
44 -7 V
43 -5 V
44 -6 V
44 -5 V
43 -5 V
44 -4 V
43 -5 V
42 -4 V
43 -3 V
42 -4 V
41 -3 V
41 -4 V
40 -3 V
40 -3 V
40 -2 V
38 -3 V
38 -3 V
37 -2 V
37 -2 V
36 -2 V
34 -2 V
34 -2 V
34 -2 V
32 -2 V
31 -1 V
30 -2 V
30 -1 V
28 -1 V
27 -2 V
26 -1 V
25 -1 V
24 -1 V
22 -1 V
22 -1 V
20 -1 V
19 -1 V
18 0 V
16 -1 V
15 0 V
14 -1 V
13 0 V
11 -1 V
10 0 V
8 -1 V
7 0 V
6 0 V
5 0 V
3 0 V
2 0 V
LT1
3114 1846 M
180 0 V
600 1077 M
70 0 V
75 0 V
2 0 V
3 0 V
4 0 V
6 0 V
7 0 V
8 0 V
9 0 V
11 0 V
12 0 V
13 0 V
14 0 V
16 0 V
17 0 V
18 0 V
19 0 V
20 0 V
22 0 V
22 -34 V
24 -48 V
24 -44 V
26 -41 V
27 -37 V
28 -35 V
28 -32 V
30 -30 V
31 -27 V
31 -26 V
32 -23 V
33 -22 V
34 -20 V
35 -19 V
35 -17 V
36 -16 V
37 -15 V
37 -15 V
38 -13 V
38 -12 V
39 -12 V
39 -10 V
40 -10 V
40 -10 V
41 -9 V
40 -8 V
41 -8 V
42 -8 V
41 -7 V
42 -6 V
41 -6 V
42 -6 V
42 -6 V
42 -5 V
41 -5 V
42 -5 V
41 -4 V
42 -4 V
41 -4 V
40 -4 V
41 -4 V
40 -3 V
40 -3 V
39 -4 V
39 -2 V
38 -3 V
38 -3 V
37 -2 V
37 -3 V
36 -2 V
35 -2 V
35 -2 V
34 -2 V
33 -2 V
32 -2 V
32 -1 V
30 -2 V
30 -1 V
28 -2 V
28 -1 V
27 -1 V
26 -1 V
24 -2 V
24 -1 V
23 -1 V
21 0 V
20 -1 V
19 -1 V
18 -1 V
17 0 V
16 -1 V
14 -1 V
13 0 V
12 -1 V
11 0 V
9 0 V
8 -1 V
7 0 V
6 0 V
4 0 V
3 0 V
2 0 V
stroke
grestore
end
showpage
}
\put(3054,1846){\makebox(0,0)[r]{Semi-classical}}
\put(3054,1946){\makebox(0,0)[r]{Exact}}
\put(2008,51){\makebox(0,0){$N|a|/a_{{\rm osc}}$ }}
\put(100,1180){%
\special{ps: gsave currentpoint currentpoint translate
270 rotate neg exch neg exch translate}%
\makebox(0,0)[b]{\shortstack{$N_0/N$ }}%
\special{ps: currentpoint grestore moveto}%
}
\put(3417,151){\makebox(0,0){20}}
\put(3135,151){\makebox(0,0){18}}
\put(2854,151){\makebox(0,0){16}}
\put(2572,151){\makebox(0,0){14}}
\put(2290,151){\makebox(0,0){12}}
\put(2009,151){\makebox(0,0){10}}
\put(1727,151){\makebox(0,0){8}}
\put(1445,151){\makebox(0,0){6}}
\put(1163,151){\makebox(0,0){4}}
\put(882,151){\makebox(0,0){2}}
\put(600,151){\makebox(0,0){0}}
\put(540,2109){\makebox(0,0)[r]{1.5}}
\put(540,1903){\makebox(0,0)[r]{1.4}}
\put(540,1696){\makebox(0,0)[r]{1.3}}
\put(540,1490){\makebox(0,0)[r]{1.2}}
\put(540,1283){\makebox(0,0)[r]{1.1}}
\put(540,1077){\makebox(0,0)[r]{1}}
\put(540,870){\makebox(0,0)[r]{0.9}}
\put(540,664){\makebox(0,0)[r]{0.8}}
\put(540,457){\makebox(0,0)[r]{0.7}}
\put(540,251){\makebox(0,0)[r]{0.6}}
\end{picture}
    \caption{Population of the lowest single-particle level 
 as a function of the effective coupling.  The exact result 
is for $N=50$ particles.}
    \label{fig:fig2}
\end{figure}

We now proceed by writing out the Hamiltonian in Eq. (\ref{eq:eq3}) 
in more detail.  
The single-particle matrix elements are given by 
\begin{equation}
 \langle 0|\hat{t}+\hat{u}|0\rangle=\frac{3}{4}\hbar\omega 
\left(x^2+\frac{1}{x^2}\right) 
\label{eq:eq5}
\end{equation}
\begin{equation}
 \langle 1|\hat{t}+\hat{u}|1\rangle =\frac{5}{4}\hbar\omega
\left(x^2+\frac{1}{x^2}\right) 
\label{eq:eq6}
\end{equation}
where $x\equiv b/a_{{\rm osc}}$, and $a_{{\rm osc}}^2=\hbar/m\omega$.  
Since $\psi_0$ and $\psi_1$ are real and of opposite parity, 
the only non-zero two-particle matrix elements are 
\begin{eqnarray}
 V_{0000}&=&-|U_0| \\
 V_{1111}&=&-\frac{3}{4}|U_0| \\ 
 V_{0011}&=&-\frac{1}{2}|U_0|=V_{0101}=V_{0110}=V_{1001}
=V_{1100} 
\label{eq:eq7}
\end{eqnarray}
where 
\begin{equation}
|U_0|=\hbar\omega\sqrt{\frac{2}{\pi}}\frac{|a|}{a_{{\rm osc}}}\frac{1}{x^3}.
\label{eq:eq8}
\end{equation}
By defining $2\epsilon\equiv\hbar\omega(x^2+1/x^2)$,   
we obtain  
\begin{eqnarray}
\hat{H}&=& \frac{1}{2}\epsilon(\hat{a}_1^{\dagger}\hat{a}_1-\hat{a}_0^{\dagger}\hat{a}_0)+2\epsilon(\hat{a}_0^{\dagger}\hat{a}_0+\hat{a}_1^{\dagger}
\hat{a}_1) \nonumber \\  
&-&\frac{1}{2}|U_0|\hat{a}_0^{\dagger}\hat{a}_0^{\dagger}\hat{a}_0\hat{a}_0 
 -\frac{3}{8}|U_0| 
\hat{a}_1^{\dagger}\hat{a}_1^{\dagger}\hat{a}_1\hat{a}_1  \nonumber \\ 
 &-&\frac{1}{4}|U_0|(\hat{a}_0^{\dagger}\hat{a}_0^{\dagger}\hat{a}_1\hat{a}_1 
 +\hat{a}_1^{\dagger}\hat{a}_1^{\dagger}\hat{a}_0\hat{a}_0) \nonumber \\ 
 &-&\frac{1}{2}|U_0|
(\hat{a}_0^{\dagger}\hat{a}_{1}^{\dagger}\hat{a}_{1}\hat{a}_0+\hat{a}_1^{\dagger}\hat{a}_0^{\dagger}
\hat{a}_0\hat{a}_1), 
\label{eq:eq9} 
\end{eqnarray}
which is of the same form as the Hamiltonian for the Lipkin model 
\cite{lip65}.  
Introducing the so-called quasi-spin operators 
\begin{equation}
 \hat{J}_z=\frac{1}{2}(\hat{a}_1^{\dagger}\hat{a}_1-\hat{a}_0^
{\dagger}\hat{a}_0)
\label{eq:eq101}
\end{equation}
\begin{equation}
 \hat{J}_+=\hat{a}_1^{\dagger}\hat{a}_0, 
\label{eq:eq102}
\end{equation}
and
\begin{equation}
\hat{J}_-=\hat{a}_0^{\dagger}\hat{a}_1,  
\label{eq:eq103}
\end{equation} 
which obey the $SU(2)$ algebra of ordinary spin, that is 
$[\hat{J}_+,\hat{J}_-]=2\hat{J}_z$ and $[\hat{J}_z,\hat{J}_{\pm}]=
\pm \hat{J}_{\pm}$, the Hamiltonian 
can be written as 
\begin{eqnarray}
\hat{H}&=&\epsilon \hat{J}_z +2\epsilon\hat{N}-\frac{1}{2}|U_0|
\left[\frac{\hat{N}}{2}
\left(\frac{\hat{N}}{2}
-1\right)-(\hat{N}-1)\hat{J}_z+\hat{J}_z^2\right] \nonumber \\ 
 &-& \frac{3}{8}|U_0|\left[\frac{\hat{N}}{2}\left(\frac{\hat{N}}{2}
-1\right)+(\hat{N}-1)\hat{J}_z+\hat{J}_z^2\right] \nonumber \\ 
 &-&\frac{1}{4}|U_0|\left(\hat{J}_+\hat{J}_++\hat{J}_-\hat{J}_-\right)
\nonumber \\
 &-&\frac{1}{2}|U_0| (\hat{J}_+\hat{J}_-+\hat{J}_-\hat{J}_+
-\hat{N}) 
\label{eq:eq11}
\end{eqnarray}
where $\hat{N}=\hat{a}_0^{\dagger}\hat{a}_0+\hat{a}_1^{\dagger}\hat{a}_1$ 
is the particle number 
operator.  The dimensionless parameter in the problem is seen to 
be $N|U_0|/\epsilon$.  
Furthermore, we introduce the ``Cartesian'' components 
of the quasi-spin through $\hat{J}_{\pm}=\hat{J}_x\pm i\hat{J}_y$ and  
rewrite $\hat{H}$ as  
\begin{eqnarray}
\hat{H}&=&\epsilon \hat{J}_z+2\epsilon\hat{N}-\frac{1}{2}|U_0|
\left[\frac{\hat{N}}{2}
\left(\frac{\hat{N}}{2}
-1\right)-(\hat{N}-1)\hat{J}_z+\hat{J}_z^2\right] \nonumber \\ 
 &-& \frac{3}{8}|U_0|\left[\frac{\hat{N}}{2}\left(\frac{\hat{N}}{2}
-1\right)+(\hat{N}-1)\hat{J}_z+\hat{J}_z^2\right] \nonumber \\ 
 &-&\frac{1}{2}|U_0|(\hat{J}_x^2-\hat{J}_y^2)-|U_0|
\left(\hat{J}_x^2+\hat{J}_y^2-\frac{\hat{N}}{2}\right). 
\label{eq:eq12} 
\end{eqnarray}
Since the total spin $\hat{J}^2=(\hat{J}_+\hat{J}_-+\hat{J}_-\hat{J}_+)/2+
\hat{J}_z^2$ commutes with $\hat{H}$,   
the eigenstates and eigenvalues can be obtained exactly by 
diagonalizing $(2J+1)\times(2J+1)$ matrices.  For $N$ particles, 
the ground state is found among configurations having $J=N/2$.  
We can also solve the problem analytically  
in the semi-classical approximation.  In this approach the 
angular momentum operators become c-numbers and we take 
\begin{equation}
 J_x=\frac{N}{2}\sin\theta \cos\phi,
\label{eq:eq131}
\end{equation}  
\begin{equation}
 J_y=\frac{N}{2}\sin\theta \sin\phi,
\label{eq:eq132}
\end{equation}
and  
\begin{equation}
 J_z=\frac{N}{2}\cos\theta. 
\label{eq:eq133} 
\end{equation}   
Upon substituting these expressions in $\hat{H}$, and 
taking $N/2(N/2-1)\approx N^2/4$, $N-1\approx N$, we find the 
semi-classical expression for the energy of the system 
\begin{eqnarray}
 E&=&\frac{N\epsilon}{2}\cos\theta+2\epsilon N -\frac{1}{2}|U_0|
[\frac{7N^2}{16}-N \nonumber \\
 &-&\frac{N^2}{8}\cos\theta+
\frac{7N^2}{16}\cos^2\theta  
 +\frac{N^2}{4}\sin^2\theta(2+\cos 2\phi)]. 
\label{eq:eq14}
\end{eqnarray}
This quantity is for fixed $\theta$ clearly minimized by $\phi=0$, 
and thus the energy per particle can be written as 
\begin{eqnarray}
\frac{E}{N}&=&\frac{1}{2}\left(\epsilon+\frac{N|U_0|}{8}\right)\eta 
+\frac{5N|U_0|}{32}\eta^2 \nonumber \\ 
 &-&\frac{1}{2}\left(\frac{19N}{16}-1\right)|U_0|+2\epsilon 
\label{eq:eq15}
\end{eqnarray}
with $\eta\equiv \cos\theta$.  
Equilibrium occurs at the value $\eta_0$ of $\eta$ given by 
\begin{equation}
   \eta_0=
    \left\{\begin{array}{cc}-1,&\frac{N|U_0|}{\epsilon}<2,\\ 
               &      \\
    -\left(\frac{1}{5}+\frac{8\epsilon}{5N|U_0|}\right) ,&
   \frac{N|U_0|}{\epsilon}\ge 2,\end{array}\right.
    \label{eq:16}
\end{equation}
and the equilibrium energy is found to be 
\begin{equation}
   \frac{E_0}{N}=
    \left\{\begin{array}{cc}-\frac{1}{2}[-3\epsilon+(N-1)|U_0|],
&\frac{N|U_0|}{\epsilon}<2,\\
    &    \\
   -\frac{1}{10}[-19\epsilon+\frac{4\epsilon^2}{N|U_0|}
+(6N-5)|U_0|] ,&\frac{N|U_0|}{\epsilon}\ge 2.\end{array}\right.
    \label{eq:eq17}
\end{equation}
For the population of the lowest single-particle state $\psi_0$ 
we get 
\begin{equation}
   \frac{N_0}{N} =
    \left\{\begin{array}{cc}1,&\frac{N|U_0|}{\epsilon}\le 2,\\
      &        \\    
     \frac{3}{5}+\frac{4\epsilon}{5N|U_0|},&\frac{N|U_0|}{\epsilon}
 > 2,\end{array}\right.
\label{eq:eq18}
\end{equation}
Thus we see that the extent to which the condensate is fragmented 
is determined by the ratio 
\begin{equation}
\frac{N|U_0|}{\epsilon}=\sqrt{\frac{8}{\pi}}\frac{1}{x^5+x}
\frac{N|a|}{a_{{\rm osc}}}.  
\label{eq:eq19}
\end{equation}
The quantity $N|a|/a_{{\rm osc}}$ will from here on be called the effective 
coupling.  A large effective coupling will give a correspondingly 
large population of the state $\psi_1$.  
However, since the two-body interaction is attractive the system 
will collapse if the effective coupling exceeds a critical value.  
From a variational calculation with only the state $\psi_0$ 
taken into account, the critical value is found to be 
$N|a|/a_{{\rm osc}}\approx 0.67$ \cite{fetter97,petsmi97}, in reasonable   
agreement with the stability criterion $N|a|/a_{{\rm osc}}<0.58$  
obtained by Ruprecht {\it et al.} from 
numerical solutions of the time-dependent Gross-Pitaevskii 
equation \cite{rupr95}.  
In Fig. \ref{fig:fig1} we show results for both the exact 
and the semi-classical calculation of the ground state 
energy of the Hamiltonian in Eq. (\ref{eq:eq11}).  
The exact solution was obtained numerically.  
The quantity $\hat{H}/N$ depends on $N$ essentially only through 
the effective coupling $N|a|/a_{{\rm osc}}$, and we checked that 
it was sufficient to do the calculations with $N=50$ particles.  
This made numerical calculations easy, as we simply had to 
diagonalize $51\times 51$ matrices for various values of the 
variational parameter $x=b/a_{{\rm osc}}$.
From Fig. \ref{fig:fig1} it is seen that the semi-classical 
solution is very close to the exact one.  
The critical 
value of the effective coupling was again found to be $\approx 0.67$.  
At this low effective coupling, the interaction is not strong 
enough to excite particles into the state $\psi_1$ to any 
appreciable extent, and the result is therefore nearly identical 
to the calculation of Ref. \cite{petsmi97}.  
This is reflected in the occupation number for $\psi_0$, shown 
as a function of $N|a|/a_{{\rm osc}}$ in Fig. \ref{fig:fig2}.  
For all effective couplings which allow for a metastable Bose 
condensate, all particles condense in $\psi_0$.  
The depletion reaches $\sim$ 10\% at $N|a|/a_{{\rm osc}}\approx 3.5$, 
and approaches 40\% in the limit $N|a|/a_{{\rm osc}} \gg 1$. 
The derivative of the semi-classical result for $N_0/N$ is discontinuous at 
$N|U_0|/\epsilon=2$, while from Fig. \ref{fig:fig2} this is  
not the case for the exact result.  This is 
due to the finite value of $N$ in the exact calculation, as 
we checked that the edge at $N|U_0|/\epsilon=2$ became sharper 
when we increased $N$.  
 
Our choice of single-particle wave functions led to specific 
ratios of the interaction strengths $V_{0000}$, $V_{1111}$,  
and $V_{1100}$.  As one could imagine alternative choices for 
the wave functions (experimentally one could alter the trapping 
potential), it is worthwhile to check how a change in these 
ratios affects our results.  We use the same form of the Hamiltonian 
in Eq. (\ref{eq:eq9}), but take 
\begin{equation}
 V_{0000}=-|U_0|,   
\label{eq:eq201}
\end{equation}
\begin{equation}
 V_{1111}=-\alpha|U_0|,  
\label{eq:eq202}
\end{equation}
and 
\begin{equation}
 V_{1100}=-\beta|U_0|. 
\label{eq:eq203}
\end{equation}
In the semi-classical approximation the energy per particle 
becomes 
\begin{eqnarray}
\frac{E}{N}&=&\frac{1}{2}\left[\epsilon+\frac{(1-\alpha)N|U_0|}{2}
\right]\eta+\frac{N|U_0|}{8}(6\beta-\alpha-1)\eta^2  \nonumber \\   
&-&(2+2\alpha+3\beta)\frac{N|U_0|}{4}+2\epsilon  
\label{eq:eq21}
\end{eqnarray}
neglecting constant terms of order 1 compared with $N$.    
Minimizing with respect to $\eta$ we find that the population 
of the lowest single-particle state is given by 
\begin{equation}
\frac{N_0}{N}=\frac{3\beta-\alpha}{6\beta-\alpha-1}
+\frac{\beta}{6\beta-\alpha-1}\frac{\epsilon}{N|U_0|}  
\label{eq:eq22}
\end{equation}
and we see that we could get a significant fragmentation of 
the condensate even with $N|U_0|/\epsilon\sim 1$ if 
we could make $\beta$ large.  
However, the elementary inequality 
\begin{equation}
\int d^3 r |\psi_0|^4 +\int d^3 r |\psi_1|^4 \ge 2\int d^3 r |\psi_0|^2 
|\psi_1|^2,  
\label{eq:eq23}
\end{equation}
which gives $1+\alpha \ge 2\beta$, 
will in practical situations prohibit us from designing the 
two-body matrix elements so that significant fragmentation 
is obtained.  If $1+\alpha=2\beta$ then $|\psi_0|^2=|\psi_1|^2$, 
i.e. $\psi_0$ and $\psi_1$ are identical and there is no 
fragmentation. 

To conclude, we have examined a simple two-level model 
to see whether an attractive interaction favors 
the formation of a fragmented Bose-Einstein condensate.  
Our results indicate that the answer to this question is 
no: at the coupling strengths where a metastable state 
exists, all particles condense in the state $\psi_0$.     
We have only considered the case of condensation in an isotropic 
harmonic oscillator potential, and the situation could be different 
if one were free to design the relative strengths of the 
two-body matrix elements involved in the calculation.  
However, the inequality (\ref{eq:eq23}) must be satisfied, 
and this seems to prevent us from choosing the strengths so 
that fragmentation is obtained.   
Thus the conclusion remains the same as in the homogeneous 
case considered by Nozi\`{e}res and St. James \cite{noz82}: 
the system will collapse before a fragmented Bose condensate 
can form.  We remark that our results are consistent with 
those of Wilkin {\it et al.} \cite{wilkin98}, who found no fragmentation 
for bosons with weakly attractive interactions in the absence 
of rotation. 

\O E is grateful for the hospitality of Nordita, where this work 
was carried out.

\end{multicols}

\clearpage

\end{document}